\documentclass[aps,pra,twocolumn,preprintnumbers,nofootinbib,amsmath,amssymb,floatfix]{revtex4-1}
\usepackage[utf8]{inputenc}
\usepackage{graphicx}
\usepackage{dcolumn}
\usepackage{bm}
\usepackage[english]{babel}
\usepackage{amsmath}
\usepackage{amssymb}
\usepackage{amstext}
\usepackage{color}
\newcommand{\beq}{\begin{equation}}
\newcommand{\eeq}{\end{equation}}
\newcommand{\eeqa}{\end{eqnarray}}
\newcommand{\beqa}{\begin{eqnarray}}
\newcommand{\ket}[1]{\left|#1\right\rangle} 
\newcommand{\bra}[1]{\left\langle#1\right|} 

\def\vec#1{{\boldsymbol{#1}}}  
\newcommand{\bs}{\boldsymbol}

\begin{document}

\title{Nanoplasmonic planar traps - a tool for engineering $p$-wave interactions}

\author{B. Juli\'a-D\'{\i}az$^{1,2}$,  T. Gra\ss$^2$, O. Dutta$^2$, D. E. Chang$^2$, and M. Lewenstein$^{2,3}$}

\affiliation{$^1$ Departament d'Estructura i Constituents de la Mat\`{e}ria,
Universitat de Barcelona, 08028 Barcelona, Spain}
\affiliation{$^2$ICFO-Institut de Ci\`encies Fot\`oniques, Parc Mediterrani 
de la Tecnologia, 08860 Barcelona, Spain}
\affiliation{$^3$ICREA-Instituci\'o Catalana de Recerca i Estudis Avan\c cats, 
08010 Barcelona, Spain}

\maketitle

{\bf 
Engineering strong $p$-wave interactions between fermions is one 
of the challenges in modern quantum physics. Such interactions are 
responsible for a plethora of fascinating quantum phenomena such as topological quantum 
liquids and exotic superconductors. In this letter we propose to combine recent 
developments of nanoplasmonics with the progress in realizing laser-induced 
gauge fields. Nanoplasmonics allows for strong confinement 
leading to a geometric resonance in the atom-atom scattering. In combination 
with the laser-coupling of the atomic states, this is shown to result in the desired 
interaction. We illustrate how this scheme can be used for the stabilization 
of strongly correlated fractional quantum Hall states in ultracold fermionic gases. 
}

\vspace{1cm}

Recently there has been growing interest in plasmonic nanostructures
that can be used for various applications in quantum optics and 
atomic physics~\cite{murphy09,chang09,stehle11,nanop,dechang07,aki07,kolesov}. 
Particularly interesting is the possibility of confining atomic 
motion over regions in space of order of nanometers, comparable 
or smaller than typical values of the atom-atom scattering length. 
In such a regime, atomic scattering undergoes strong modifications 
due to {\it confinement induced resonances}~\cite{olshanii, petrov}. 
Here we propose to use this effect to engineer strong and robust 
$p$-wave interactions between fermionic atoms in planar geometries, 
which overcomes the challenges associated with creating such 
interactions in previously proposed techniques. This opens a new 
path towards the realization of exotic fractional 
quantum Hall states~\cite{lau,MR91} and superfluid 
phases~\cite{Readgreen,Read}.  

A strong motivation for realizing such states are their intriguing 
topological properties which find direct application in topological 
quantum computation, protected quantum qubits, and protected quantum 
memories~\cite{NS08}. Similarly, $p$-wave repulsion can stabilize 
low filling fractional quantum Hall states~\cite{lau,regfer}, including the Moore-Read 
state~\cite{MR91}. This state has been proposed in 
the context of a pronounced fractional quantum Hall plateau at filling 
5/2~\cite{Read}, but formally it also resembles the spinless chiral 
$p$-wave superfluid state. In solid-state physics, only in Strontium 
Ruthenate, chiral ($p_x + i p_y$)-wave Cooper pairs are believed 
to be responsible for the observed superfluidity of 
electrons~\cite{Rice}. In the field of quantum gases, strong 
$p$-wave interaction can in principle be achieved by using Feshbach 
resonances. Due to the inelastic loss processes, however, a strong 
$p$-wave interaction is hard to achieve experimentally~\cite{LC08}. 
Also, in Bose-Fermi mixtures, density fluctuations of bosons can 
induce attractive $p$-wave interactions or even higher partial waves 
between the fermions~\cite{EV02, OM,pietro}. However, such proposals also run into 
difficulties due to the phase separation instability of 
Bose-Fermi mixtures and stringent constraints on temperature.

\begin{figure*}[t]
\vspace{10pt}
\includegraphics[width=1.\textwidth,angle=-0]{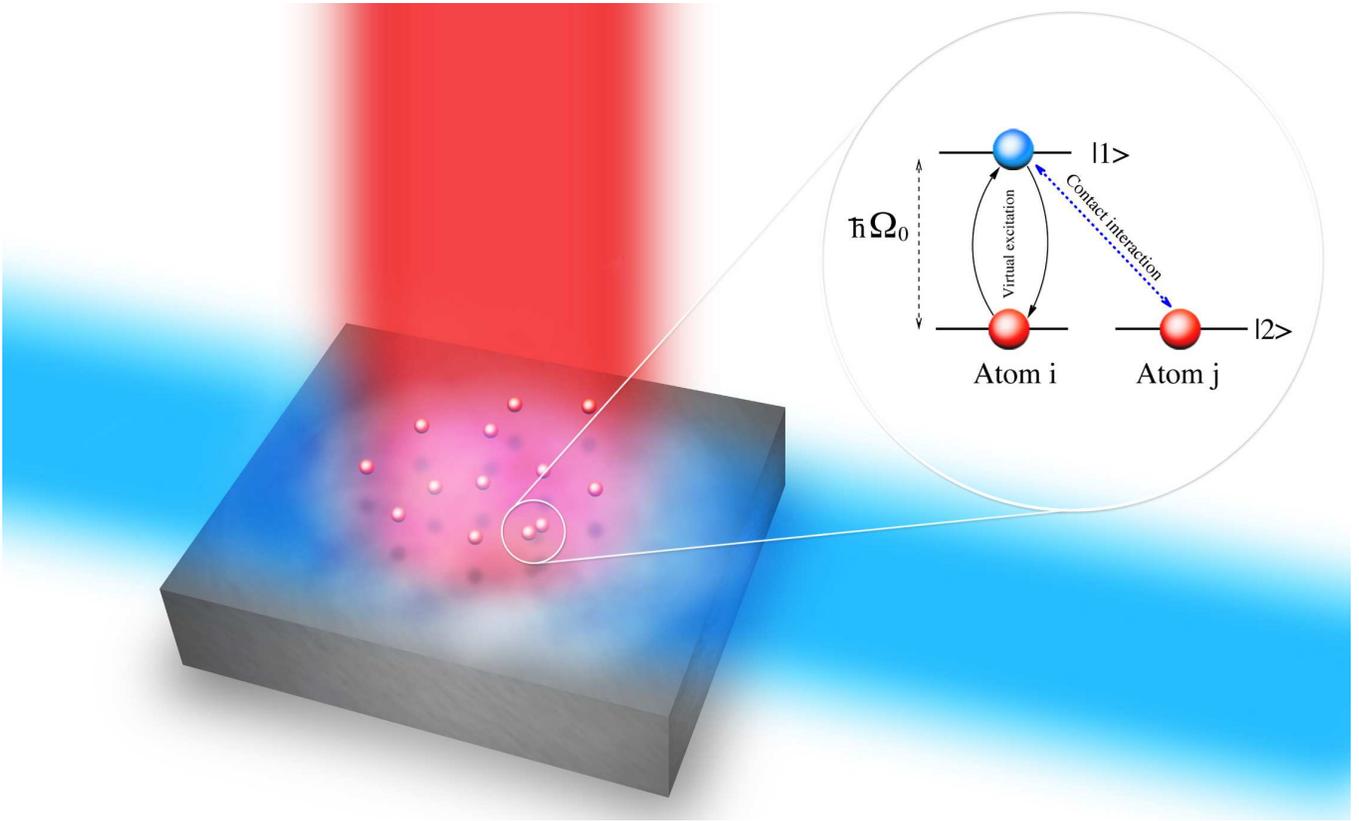}
\caption{\label{nicefig} 
The ultracold atomic sample is tightly confined above a 
metallic surface by a nanoplasmonic field produced by an 
external laser (in red) pointing perpendicular to the 
metal surface. A second laser (in blue), shining in from 
the side, is used to generate the artificial magnetic field 
felt by the atoms. The mechanism for fermion-fermion contact
interaction involves the virtual excitation of one of the atoms 
into the excited dressed state, as illustrated in the inset.}
\end{figure*}

In this letter we propose to combine two important concepts: 
strongly confined two-dimensional (2D) traps via nanoplasmonic 
fields, and strong laser induced synthetic gauge fields, as 
illustrated in Fig.~\ref{nicefig}. For simplicity, the synthetic 
gauge field considered is produced through a minimal scheme 
described in Ref.~\cite{Dalibard:2011}. It consists of a laser 
field coupling two internal levels of the fermionic atoms, 
and an external electric or magnetic field which 
produces a linear variation of the energy of the internal 
states throughout the sample. Preparing the system in the lower 
dressed state, the sample is effectively subjected to a 
strong synthetic magnetic field~\cite{dali}. Being 
polarized in one dressed state, interactions between 
the fermionic atoms are prohibited by the Pauli principle. However, 
the external degrees of freedom provide a small coupling to the 
higher dressed state. Remarkably, this will be shown to result 
in a residual $p$-wave contact interaction between the fermions. 
This contribution can be enhanced thanks to the resonant 
behavior of the atom-atom scattering length in strongly confined 
2D settings~\cite{petrov}. This not only allows to strengthen 
the interaction but also to explore both attractive and repulsive 
$p$-wave interactions, i.e. going from the physics of $p$-wave pairing 
to fractional quantum Hall physics. 

Experimental difficulties to provide a sufficient transverse confinement, 
that is on the order of the atom-atom scattering length, are surmountable 
thanks to the new developments in plasmonics. The interaction of cold 
atoms with nanoplasmonic systems has attracted significant interest 
recently. A notable feature of surface plasmon excitations, which exist 
along a metal-dielectric interface, is the lack of a diffraction limit. 
In the context of atom trapping, this enables the generation of fields 
with dramatically reduced effective wavelengths compared to free space, 
and a corresponding reduction of parameters such as trap confinement. 
The properties of the plasmons can also be greatly engineered through 
the underlying device geometry. The interaction between 
Bose-Einstein condensates and tailored plasmonic micro-potentials 
has recently been observed~\cite{stehle11}, and plasmon-based trapping 
techniques for ultracold atoms with applications in quantum 
simulation have been proposed~\cite{murphy09,chang09,nanop}. 

As a straightforward application of the scheme, we will consider 
the case of repulsive interactions and show by exact diagonalization 
how the induced $p$-wave interaction can be used to explore different 
quantum Hall phases, notably going from filled Landau level (LL) 
physics, to the fractional quantum Hall regime with a $\nu=1/3$ 
Laughlin state~\cite{lau}, passing through a phase with sizable 
overlap with the Pfaffian state~\cite{MR91}.

\subsection*{\bf Model}

We consider a trapped ultracold gas of fermionic atoms with two internal 
states $|{\rm g}\rangle, |{\rm e}\rangle$. The single particle Hamiltonian 
$H_{\rm sp}= H_{\rm ext} + H_{\rm AL}$ consists of an external part 
$H_{\rm ext}=p^2/(2M) + V({\bf r})$ with the anisotropic trapping potential 
$V({\bf r})$, and an atom-laser coupling $H_{\rm AL}$ including also the internal 
energies. This coupling is responsible for a synthetic gauge field which 
emerges due to the accumulation of Berry's geometrical phase when an atom 
moves within the laser field~\cite{berry}. The key to achieving non-vanishing 
phases on closed contours is to make the internal energies, and thus 
$H_{\rm AL}$, spatially dependent via a Stark or Zeeman shift, such that 
$H_{\rm AL}$ and $H_{\rm ext}$ do not commute (see Methods). 

The laser light mixes the ground and excited state, giving rise to 
position-dependent {\it dressed states}, $\ket{\Psi_1}$ and 
$\ket{\Psi_2}$, which are the eigenstates of the atom-laser 
interaction. As detailed in the Methods section, 
increasing the laser strength, and thus the Rabi coupling between 
the two bare states, one can energetically favor one dressed manifold, 
say $\ket{\Psi_2}$. By adjusting the external trapping, the single 
particle Hamiltonian projected in this lowest dressed manifold can 
be written as the usual quantum Hall one, 
\beqa
H_{22}
& = & \frac{({\bs p} + {\bs A})^2}{2M}
+\frac{M \omega_\perp^2}{2}(1-\eta^2) (x^2+y^2)
\label{eq9}
\eeqa
where $\omega_{\perp}$ is the effective $xy$ trapping frequency, 
${\bs A}=\hbar \eta (y,-x) /\lambda_\perp^2$, and $\eta$ is the 
strength of the synthetic gauge field which depends on the laser 
wavenumber $k$ and the spatial extent of the Stark or Zeeman 
shift $w$ (see Methods). This Hamiltonian has the well-known LL 
structure, and its eigenfunctions are the Fock-Darwin states. 
Restricting ourselves to the lowest LL, the corresponding wave 
functions read $\varphi^{\rm FD}_l(z) \propto z^l \exp(-|z|^2/\lambda_{\perp}^2)$ 
where $z=x-iy$ describes the atom position in the $(x,y)$ plane, and 
$\lambda_\perp = \sqrt{\hbar/(M\omega_\perp)}$. The adiabatic approximation 
requires large Rabi frequencies $\hbar\Omega_0\gg E_R$~\cite{dali}, 
where the recoil energy is 
$E_R= (k^2 \lambda_\perp^2/ 2) (\hbar \omega_\perp)$. 
Within this limit, the off-diagonal Hamiltonian elements, $H_{12}$ and 
$H_{21}$, connecting the dressed states 
are neglected. Then, transitions to the higher dressed manifold 
are fully suppressed. A general atomic state, 
$\chi(\vec{r})=\tilde{\varphi}_1(\vec{r}) \otimes \ket{\Psi_1}
+\tilde{\varphi}_2(\vec{r})\otimes \ket{\Psi_2}$, 
becomes a low-lying solution for $\tilde\varphi_1=0$ and 
$\tilde\varphi_2 = \varphi^{\rm FD}_l$. 
In our approach, however, some amount of non-adiabaticity is 
crucial, as it will yield a finite value for $\tilde\varphi_1$ 
resulting in non-zero contact interactions.

\subsection*{\bf $p$-wave fermion-fermion interaction}

Now we turn to the atom-atom interactions, which we take as 
contact interactions. In terms of the bare fermionic states, 
it reads
\beq
V_{ij} =g_c {\hbar^2\over M} \ \delta(z_i-z_j) 
( \ket{\rm e} \ket{\rm g} \bra{\rm e}\bra{\rm g}
+
  \ket{\rm g} \ket{\rm e} \bra{\rm g}\bra{\rm e}
 ) \,.
\label{vint}
\eeq
Here, $g_c$ is a number quantifying the interaction strength. 
A more precise definition will be given later. Of course, in 
the dressed basis the interaction term maintains its form, 
such that interactions remain restricted to pairs of one atom 
in $\ket{\Psi_1}$ and the other in $\ket{\Psi_2}$. Thus, by 
polarizing the system in the lower dressed state $\ket{\Psi_2}$, 
no interactions are present in the adiabatic limit 
$\Omega_0 \rightarrow \infty$. Still, by making the ratio of 
the Rabi frequency to recoil energy much bigger than 1, 
$R_E \equiv \hbar \Omega_0 / E_R\gg1$, we can work in a 
quasi-polarized regime, in which the $\ket{\Psi_1}$ level 
serves only as a virtual manifold.

In this limit, the unperturbed many-body Hamiltonian is given by
\begin{align}
 \label{a1}
H^{(0)}=\sum_{i=1}^N  H^i_{22} {\cal P}_i\,.
\end{align}
where the operator 
${\cal P}_i=\ket{\Psi_2}_i\bra{\Psi_2}_i$ projects the $i$th 
particle onto the low-lying Hilbert space.
%
%
%
The off-diagonal terms, $H_{12}\ket{\Psi_1}\bra{\Psi_2}$ and
$H_{21}\ket{\Psi_2}\bra{\Psi_1}$, and the atom-atom 
interaction of Eq.~(\ref{vint}) are taken as perturbations. They 
give second-order corrections. The effective many-body Hamiltonian 
can then be written as
\beqa
H^{\rm eff} &=& H^{(0)} + H^{(1)} + H^{(2)} \nonumber \rm{ \ \ with} \\
H^{(1)} &=&- \sum_i {H^i_{21} \ H^i_{12}\over \hbar\Omega_0} 
{\cal P}_i \nonumber\\
H^{(2)}&=& \sum_{ij} {\cal P}_i {H_{21}^i\  V_{ij} \ H_{12}^j \over (\hbar\Omega_0)^2} 
{\cal P}_j 
\,.
\label{eq4}
\eeqa
Note that the denominator in $H^{(1)}$ has been set to a constant 
equaling the energy difference between dressed states $\ket{\Psi_2}$ 
and $\ket{\Psi_1}$. As this is taken to be large, it is the 
dominant contribution to the energy gap. 

In a previous study of a bosonic system~\cite{dali,ournjp}, we 
have analyzed the influence of $H^{(1)}$, but the many-body 
contribution $H^{(2)}$ has been negligible due to the bosonic 
nature of the atoms. We will in the following show that in 
the fermionic case, where $H^{(2)}$ is the only many-body 
contribution, it becomes crucial. As illustrated in 
Fig.~\ref{nicefig}, $H^{(2)}$ describes a process where one 
atom is excited from $\ket{\Psi_2}$ to the virtual $\ket{\Psi_1}$ 
manifold, where it interacts with an atom in $\ket{\Psi_2}$, to then 
get de-excited to $\ket{\Psi_2}$ again. Importantly, acting among 
fermions, the many-body interaction term gives solely non-zero 
$p$-wave contributions. This is seen by using Eq.~(\ref{H12}) to 
cast $H^{(2)}$ into $p$-wave form (cf. Ref.~\cite{regfer}), 
\beq
H^{(2)} \propto \sum_{i,j} \hat p_{ij} 
\ \delta^{(2)}(z_{ij}) \hat p_{ij} \ {\cal P}_i {\cal P}_j
\label{econtact}
\eeq
with the relative variables, 
$\hat p_{ij}=-i \hbar (\partial_{z_i} -\partial_{z_j})$ and $z_{ij}=z_i-z_j$.  
The important feature of Eq.~(\ref{eq4}) is that the effective 
interaction is linear in the bare one $V_{ij}$, which allows 
one to change the interaction from attractive to repulsive. 
This is in contrast to second order mechanims like the Kohn-Luttinger~\cite{kohn}. 

The main question which arises at this point is whether 
the residual interaction term, Eq.~(\ref{econtact}), 
is strong enough to significantly modify the physics of the system. 
This becomes possible by tuning the interaction strength $g_c$. 
It is well known that this parameter crucially depends on the 
geometry of the system. In particular, for transversal 
confinements on the order of the scattering length, and 
considering the case of attractive interaction the effective 2D 
coupling is known to behave as~\cite{petrov}
\beq
g_c={4\pi \hbar^2\over M} 
{1\over 
\sqrt{2\pi}\lambda_z/ a_{3D} 
+ \log(0.918 \hbar \omega_z/\pi\epsilon)}\,,
\eeq
where $\epsilon$ is the energy of the motion in the 
$x-y$ plane and $a_{3D}$ the 3D scattering length. For 
a value of $\hbar\omega_z/\epsilon=10^3$, it produces a 
resonant behavior for values of the transverse confinement 
$\lambda_z \sim 0.4 |a_{3D}|$. This confinement-induced 
resonance behaviour is not present in usual experiments 
with optical traps. There, the trapping on the $z$ direction 
has at most been of the order of hundreds of nanometers, far 
from the resonance region. The transverse confinement lengths 
of $\lambda_z\sim 5$-$10$ nm needed to facilitate significant 
interactions can be achieved using novel plasmon-based trapping 
techniques, such as those investigated theoretically and 
experimentally in Refs.~\cite{stehle11,murphy09,chang09,nanop}. 
For example, it is possible to tailor a two-dimensional array 
of metallic nanosystems~(such as nanoshells~\cite{nanop}), which 
creates a near-planar trapping potential arising from spatial 
interference between an incident field and plasmon-enhanced 
near-field. The effective wavelength characterizing this trapping 
potential scales like the characteristic size of an individual 
nanosystem, $\lambda_{\footnotesize\textrm{eff}}\sim r$, even for 
system sizes far below the free-space wavelength $r\ll\lambda_0$. 
This yields a corresponding reduction of $\sim\sqrt{r/\lambda_0}$ in 
the trap spatial confinement compared to free-space techniques.

This resonant behavior can in principle be used to produce 
arbitrarily large values of $g_c$ and, importantly, allows to 
achieve not only large values of the coupling, but also provides 
a way of producing both attractive and repulsive $p$-wave interactions 
between the fermions.

\subsection*{\bf Example: stabilization of the $\nu=1/3$ Laughlin state}

\begin{figure}[t]
\includegraphics[width=0.49\textwidth,angle=-0]{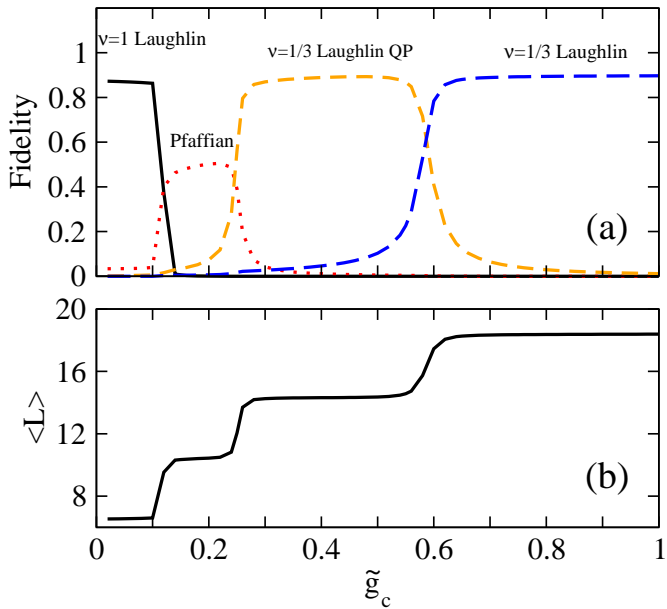}
\caption{\label{figure} 
Evolution of the ground state as a function
of the interaction parameter 
$\tilde g_c= g_c / (k \lambda_\perp \ \eta R_E)^2 $. Panel (a) 
presents the overlap of the ground state with the filled LL 
state (solid-black), the fermionic Pfaffian state (dotted-red), the 
quasiparticle state over the $\nu=1/3$ Laughlin (dashed-orange) and 
the $\nu=1/3$ Laughlin (long-dashed-blue). Panel (b) contains the 
average angular momentum of the ground state of the system. }
\end{figure}

As an example, we discuss the case of repulsive $p$-wave 
interaction. In the context of quantum Hall physics we have 
to ask whether the obtained $p$-wave interaction is capable of 
bringing the system from the integer quantum Hall regime 
of a non-interacting system to the fractional quantum Hall 
regime. In that case, a Laughlin-like state should show up 
as the ground state of the system~\footnote{Let us recall the 
Laughlin wave function~\cite{lau} at filling $\nu$, 
\beq
\Psi_{\rm Laughlin} = {\cal N} \prod_{i<j} (z_i-z_j)^{1/\nu} {\rm e}^{-|z|^2/2}\,.
\label{lau}
\eeq
which has $L_z=\nu^{-1} N(N-1)/2$.}. Note that in the quantum 
Hall regime, $\eta \rightarrow 1$, the contribution of $H_{22}$ 
reduces to a constant, as all Fock-Darwin states become 
(quasi)degenerate. Thus, to bring the system into the fractional 
quantum Hall regime, the interaction term must be comparable 
to the contribution of $H^{(1)}$ term, which breaks the 
rotational symmetry~\cite{ournjp}. 

To give definite numerical predictions, we perform an exact 
diagonalization (see Methods) with a few number of atoms, $N=4$. 
The parameters of the system are taken 
as $k=10/\lambda_\perp$, $\hbar \Omega_0=100 E_R$ and 
$\eta=0.98$. We discuss the different phases appearing as we 
vary the interaction strength, $g_c$. For weak interactions, 
the ground state of the system has a large overlap with the 
analytical form of the filled LL state, $\nu=1$, as depicted in
Fig.~\ref{figure}. The angular momentum of the ground state 
is found to be slightly larger than the analytical value, 
$L=6$. As explained in Refs.~\cite{dali,ournjp},
this is due to the derivation from rotational symmetry. When the interaction is
increased, the system undergoes a transition into a phase, where the ground
state has large overlaps with the fermionic Moore-Read state~\cite{MR91}. 
Even stronger interactions bring the system into a state which resembles the
quasiparticle excitation of the $\nu=1/3$ Laughlin state. At another 
critical value of $g_c$, one finally reaches the $\nu=1/3$ Laughlin state. 
As in the case of bosons with contact interactions, this state has zero 
interaction energy, and thus for any stronger interaction parameter, 
it remains the ground state. 

\subsection*{\bf Summary}

We have presented a novel mechanism to realize sizable $p$-wave interactions 
between fermionic atoms. The key is the combination of a strongly confining
plasmonic field, which allows to explore confinement-induced resonances, 
with a simple scheme to generate a strong synthetic gauge field. To exemplify 
the potential of our approach, we have considered the case of repulsive $p$-wave 
interaction. We have shown that our proposal allows to stabilize a number of 
interesting quantum Hall states, like the Pfaffian, and the $\nu=1/3$ Laughlin state.
In our numerical calculation we have considered a small number of atoms, as 
has become experimentally feasible recently~\cite{kino,gemelke,jocim11}, but 
we note that the scheme should also be applicable to large systems. A good 
candidate for realizing the proposal are Ytterbium atoms due to the 
long-lived state of the clock transition. Requiring an ultratight 
trapping in a 2D geometry, our proposal shall trigger the use of 
nanoplasmonic fields as a promising technique for achieving that goal.

\vspace{1cm}

\begin{acknowledgements}
BJD and TG are grateful for stimulating discussions with 
the ``Ytterbium team'' at ILP (Hamburg), C. Becker, S. D\"orscher, 
B. Hundt, and A. Thobe. This work has been supported by EU (NAMEQUAM, AQUTE),
ERC (QUAGATUA), Spanish MINCIN (FIS2008-00784 TOQATA),
Generalitat de Catalunya (2009-SGR1289),
Alexander von Humboldt Stiftung, and AAII-Hubbard.
BJD is supported by the Ram\'on y Cajal program.
ML acknowledges support from the Joachim Herz Foundation and Hamburg
University. DEC acknowledges support from Fundaci\'o Privada Cellex Barcelona.
\end{acknowledgements}

\section*{\bf Methods}

\subsection*{\bf Single particle Hamiltonian}

The single particle Hamiltonian reads, 
\beq
H_{\rm sp}= H_{\rm ext} + H_{\rm AL}
\eeq
where $H_{\rm ext}=p^2/(2M) + V({\bf r})$ with the 
anisotropic trapping potential $V({\bf r})$. $H_{\rm AL}$ is the 
atom-laser coupling, which includes the internal energies. 
To make $H_{\rm AL}$ spatially dependent, such that $H_{\rm AL}$ 
and $H_{\rm ext}$ do not commute, we perform a Stark or 
Zeeman shift of the internal energies. The strength of this shift can be 
characterized by a length scale $w$, which is chosen such 
that the energies of the bare internal states read 
$E_g= -\hbar \Omega_0 x/(2w)$ and 
$E_e=\hbar \omega_A +\hbar \Omega_0 x/(2w)$. Here, $\omega_A$ is the 
energy difference of the bare states. 
In this way, preparing the system in the ground-state of $H_{\rm AL}$, 
the external part will stimulate transitions into the excited manifold 
of $H_{\rm AL}$. The probability of such transitions is controlled by 
the Rabi frequency $\Omega_0$ of the coupling. The laser frequency is 
set to resonance with the atomic transition. Furthermore, we choose 
the laser to be a running wave in $y$-direction with wavenumber $k$. 
Then, within the rotating-wave approximation, the atom-laser
Hamiltonian $H_{\rm AL}$ can be written in terms of bare 
states $\ket{{\rm e}}$ and $\ket{{\rm g}}$ as~\cite{coh} 
\begin{align}
 \hat{H}_{\rm AL} =
\frac{\hbar \Omega}{2} \left[ \cos\theta \left(\ket{{\rm e}}\bra{{\rm e}} - \ket{{\rm g}}\bra{{\rm g}} \right) \right.\nonumber\\
\left.+ \sin\theta \left( e^{i\phi} \ket{{\rm e}}\bra{{\rm g}} + {\rm h.c.} \right)\right],
\label{eqal}
\end{align}
where $\Omega=\Omega_0\sqrt{1+x^2/w^2}$, $\tan \theta=w/x$, and $\phi=k y$. 
Note that spontaneous emission processes are not considered in the Hamiltonian 
of Eq.~(\ref{eqal}). This is justified if the two atomic states are sufficiently 
long-lived, as is the case for the $^1S_0 \rightarrow ^1P_1$ clock transition 
in Ytterbium. In contrast to the bosonic Ytterbium isotopes, the finite spin 
of the fermionic isotopes yields a small magnetic moment, which allows for 
a strong coupling of the clock states at reasonable laser power. Thus, 
achieving large Rabi frequencies, as required by our proposal, poses no 
problem for $^{171,173}$Yb~\cite{yb}.

Diagonalizing Eq.~(\ref{eqal}) yields the dressed states,
$
\ket{\Psi_1} ={\rm e}^{-iG} \left( C\,\, {\rm e}^{i\phi/2} \ket{{\rm g}} +
S\,\,{\rm e}^{-i\phi/2}  \ket{{\rm e}}\right)$, 
$\ket{\Psi_2} ={\rm e}^{iG} \left(-S\,\, {\rm e}^{i\phi/2}\ket{{\rm g}}+
C\,\,{\rm e}^{-i\phi/2} \ket{{\rm e}}\right)$, 
where $C=\cos{\theta/2}$, $S=\sin{\theta/2}$, 
$G=\frac{kxy}{4w}$. The single-particle Hamiltonian
$H_{\rm sp}$ can be expressed as a 2$\times$2 matrix $H_{ij}$. In the 
dressed state basis, its diagonal terms can be written as~\cite{dali}, 
\beq
H_{jj}=\frac{\left(\bs p-\epsilon_j \bs A \right)^2}{2M} 
+ U+V +\epsilon_j\frac{\hbar \Omega}{2}\,,
\label{eq14} 
\eeq 
with $\epsilon_1=1$ and $\epsilon_2=-1$. Full expressions for the 
vector potential $\bs A$ and the scalar potential $U$ are 
given in Ref.~\cite{dali}. Note that for $w \ll x,y$, we 
recover the symmetric gauge expression 
$\bs A(\bs r)=\frac{\hbar  k}{4w}(y,-x)$. With a convenient 
choice of the trapping potential, $H_{22}$ can be made 
symmetric. Then, the Hamiltonian element $H_{22}$ reads
\beqa
H_{22}
& = & \frac{({\bs p} + {\bs A})^2}{2M}
+\frac{M \omega_\perp^2}{2}(1-\eta^2) r^2
\label{eq9}
\eeqa
where $\omega_{\perp}$ is the effective $xy$ trapping frequency, 
$\eta =(k\lambda_\perp^2)/(4w)$, and $\lambda_\perp=\sqrt{\hbar/(M\omega_\perp)}$. 
The recoil energy of the atoms is defined as 
$E_R= (k^2 \lambda_\perp^2/ 2) (\hbar \omega_\perp)$. 

Retaining up to quadratic terms, the off-diagonal Hamiltonian 
elements, $H_{12}=H_{21}^\dagger$, explicitly read
\beqa
H_{12}&\simeq&
-\frac{\hbar ^2}{2 M}
\left[ -i k \partial_y\Psi 
+
\left(\frac{k^2 x}{4 w} +\frac{iky}{4w^2}\right)\Psi 
+
\frac{1}{w}  \partial_x\Psi \right] \nonumber\\
&=&
-\frac{\hbar ^2}{4 M} 
\left[ 
        \hat{a} c_1
+\hat{a}^\dagger c_2
+       \hat{b} c_3
+\hat{b}^\dagger c_4
\right]
\label{eq:h12r}
\,,
\eeqa
with 
$\hat{a}^\dagger\equiv -\lambda_\perp \partial_{\bar{z}} +\lambda_\perp^{-1} {1\over 2}z$, 
$\hat{a} \equiv \lambda_\perp \partial_{z} +\lambda_\perp^{-1} {1\over 2}\bar{z}$, 
$\hat{b}^\dagger\equiv -\lambda_\perp \partial_{z} +\lambda_\perp^{-1} {1\over 2}\bar{z}$, and 
$\hat{b} \equiv \lambda_\perp \partial_{\bar{z}} +\lambda_\perp^{-1}{1\over 2}z$.
Acting on a Fock-Darwin state the operators $\hat{a}$ ($\hat{a}^\dagger$) 
decrease (increase) the $l$ quantum number by one, while the operators 
$\hat{b}$ and $\hat{b}^\dagger$ change the Landau level.

As we will be interested in the fractional quantum Hall regime of 
large synthetic magnetic field, $\eta\simeq 1$, it is possible to 
safely neglect the $\hat{b}$ and $\hat{b}^\dagger$ contributions. In this 
limit, we have $c_1=c_2\simeq 8 w / \lambda_\perp^3$, and 
$c_3=-c_4\simeq 2 /( w\lambda_\perp)$. In our numerics, we will 
furthermore choose $w\simeq 2.5 \lambda_\perp$ and $k=10/\lambda_\perp$, 
implying $\eta\simeq 1$, guaranteeing $c_1 \ll c_3$. We can then write
\beqa
H_{12}
&=&
-{2 w \hbar^2\over M \lambda_\perp^3}(\hat{a}+\hat{a}^\dagger) 
+ {\cal O}[(w/\lambda_\perp)^{-2}].
\label{H12}
\eeqa

\subsection*{\bf Exact diagonalization}

To solve the effective Hamiltonian, we perform exact diagonalization. 
Therefore, we build many-body states using as single particle states 
the Fock-Darwin states, $\ket{l}$. Then the second quantized form of 
$H^{(2)}$ is
\beq
H^{(2)} = \frac{1}{2}
\sum_{ij,kl} \hat{f}^\dagger_i \hat{f}^\dagger_j \hat{f}_k
\hat{f}_l \ 
V_{ij,kl} \,,
\label{contact}
\eeq
where $\hat{f}_i$ anihilates an atom in $\varphi^{\rm FD}_i(z)$. 
The matrix element reads, 
$
\nobreak{V_{ij,kl} = (\hbar \Omega_0)^{-2}
\langle i| \langle \,j| H_{21} V H_{12} | l\rangle |k\rangle\,.}
$
Taking into account the Pauli principle, we get, 
\beqa
V_{ij,kl}^{(2)} &=& g_c \ \hbar \omega_\perp \bigg[
(\varphi^*_i  h^*_j \varphi_k  h_l ) 
-(\varphi^*_i  h^*_j h_k \varphi_l) \nonumber\\
&&-(h^*_i  \varphi^*_j \varphi_k  h_l)
+(h^*_i  \varphi^*_j h_k \varphi_l)\bigg] \,,
\label{vsuper}
\eeqa
where $h_l \equiv [H_{12}/(\hbar\Omega_0)]\  \varphi^{\rm FD}_l$.  
With the expression for $H_{12}$ from Eq.~(\ref{H12}), $h_l$ is 
directly found to be
$
h_l = (k \lambda_\perp \ \eta R_E)^{-1}
(\sqrt{l+1} \ \varphi_{l+1} +\sqrt{l} \ \varphi_{l-1})\,.
$
It is worth noting that due to the contact nature of the interaction, 
Eq.~(\ref{contact}) commutes with $\hat{L}_z$.

\bibliographystyle{apsrev4-1}

\end{document}